# Library based identification and characterisation of polymers with nano-FTIR and IR-sSNOM imaging

Michaela Meyns,* Sebastian Primpke, Gunnar Gerdts

Alfred-Wegener-Institute, Helmholtz Centre for Polar and Marine Research, Biologische Anstalt Helgoland, Kurpromenade 201, 27498 Helgoland, Germany

*Email: michaela.meyns@awi.de

*Abstract*

*AFM is a technique widely applied in the nanoscale characterisation of polymers and their surface properties. With nano-FTIR and IR-sSNOM imaging an optical dimension is added to this technique that allows for straightforward high resolution characterisation and spectroscopy of polymers. As the volume sampled by these near-field techniques depends mostly on the radius of the cantilever tip, typically 10 nm, it is orders of magnitude smaller than in conventional techniques. Nevertheless, comparability of nano-FTIR near-field spectra and data from macroscopic methods has been shown. Some relevant polymers such as polystyrene however, prove to be more difficult to detect than others. Furthermore, the small sampled volume suggests lower signal quality of nano-FTIR data and proof of its suitability for a reliable library search identification is lacking. To evaluate the techniques especially towards automatic and higher throughput identification of nanoscale polymers, for example in blends or environmental samples, we examined domain distributions in a PS-LDPE film and spectral responses of foils of the most relevant commercial polymers. We demonstrate the successful library search identification of all samples with nano-FTIR data measured in less than seven minutes/spectrum. We discuss aspects affecting the accuracy of the identification and show that already the small spectral range of 1700-1300 $cm^{-1}$ leads to similar success in differentiating between polymer types with near-field data as with conventional far-field FTIR spectroscopy.*

Fourier-Transform Infrared (FTIR) spectroscopy nowadays is a common tool for polymer characterisation and identification in research and industrial contexts. Macroscopic film samples are quickly analysed by attenuated total reflection (ATR)-IR. With this technique, the samples are pressed to the surface of a crystal and the spectrum, the chemical fingerprint of the material, is derived from the reflection of the infrared beam at the interface. By easy steps, the spectra are then compared with a spectrum library and within minutes the identification is complete. The limit of resolution is set by the diffraction limit of light.

A novel technique that moves the spatial resolution of IR-spectroscopy from micrometres in ATR to the nanometre range is nano-FTIR. The combination between a scattering-type near-field optical microscope (s-SNOM) and a mid-infrared continuum laser joins the local resolution of AFM with chemical identification by infrared spectroscopy. By directing the laser beam to the metallic AFM tip a nanofocus is created,[1] whose diameter is only limited by the dimension of the tip. Thus, a local spectroscopic resolution of up to 20 nm is possible.[2] Within this focus, near-field (NF) interactions between probe and sample occur, which are controlled by the wavenumber ($\omega$)-dependent dielectric function $\varepsilon(\omega)$ of the sample and thus its absorption/reflection properties.[2-4] Measurements are conducted in AFM tapping mode, so that the intensity of sample and background signal vary with the motion of the tip. The optical signal is recorded by interferometric detection in an asymmetric Michelson configuration, which differs from conventional interferometers as sample and tip replace the stationary mirror at the end of one of the four arms. In the adjacent arm a second mirror controlled by a piezo drive moves along the axis of the beam in the arm to provide interferograms in the form of the intensity of the optical signal as a function of distance. The ratio of light backscattered at the tip-sample system to the incident light is described by the complex valued scattering coefficient $\sigma(\omega) = s_n(\omega)e^{i\varphi_n(\omega)}$ with NF-amplitude $s_n$ and NF-phase $\varphi_n$. With higher orders $n$ of the tapping frequency $\Omega$ of the metallic cantilever tip, background interference is reduced. Fourier transformation of the detected optical signal at $n \geq 2$ results in NF-amplitude and NF-phase spectra, which translate to the local reflection and absorption of the sample, respectively.[5] The mirror of the interferometric detection unit can be adjusted to the white light position (WLP), where both optical paths have the same length and the intensity of the interferogram is at its maximum. This setting does not provide spectral resolution but a 2D NF-amplitude scan identifies regions with different reflective properties in the spectral region of the incident beam. In a next step, point spectra of different spots on the sample and a reference spectrum are obtained by Fourier transformation of the corresponding interferograms with a moving mirror. The reference is usually measured on a reflective and non-absorbing, sample-free spot on the substrate or clean Si or metal surfaces. Before further analysis, the spectra have to be normalised to the reference by dividing the obtained amplitudes $s_n/s_{n,ref}$ and subtracting the phase of the reference $\varphi_n - \varphi_{n,ref}$. For comparison of near-field with conventional far-field FTIR-spectra, it has been shown that the nano-FTIR absorption $a_n(\omega)$, the imaginary (Im) part of the $a_n(\omega) = Im[\sigma(\omega)] = s(\omega)\sin[\varphi_n(\omega)]$, may correlate better to the reference data than only $\varphi_n$,[2, 6,7] depending on the thickness of the sample.[8]

An attractive way to examine material distributions and interfaces of samples with known components is IR-sSNOM imaging, where a monochromatic source is coupled in for the optical trace. In this case, excitation close to absorption bands of the components can be employed to differentiate between materials with high resolution, e.g. 40 nm for a PMMA-PC sample.[9] The reflected beam undergoes pseudoheterodyne detection with successive deconvolution at $n \geq 2$ for $\Omega$.[5] Analogous to nano-FTIR, optical amplitude and phase intensities indicate regions of reflection and absorption of the now monochromatic radiation. Both amplitude and phase are recorded on separate channels, so that 2D scan data of both is collected in addition to the usual topographic and phase maps derived from AFM scans. On a standard polymer blend sample, we here combine nano-FTIR and IR-sSNOM-imaging and show phase distributions and spectral transitions between the domains in a line scan.

A few studies reported near-field point spectroscopy and hyperspectral imaging, where a 3D Data cube with spectra taken for each pixel is obtained.[2, 10] However, a correlation of near-field spectroscopy with a library spectrum search routine, as for example applied in the identification of unknown samples, is lacking. Furthermore, the widely distributed polymers polyethylene and especially polystyrene are more difficult to detect by near-field spectroscopy than others are.[10] For this reason, we here examine the applicability of near-field spectra with minimum post-treatment in a library-based identification of a selection of relevant polymer samples.

**Results and discussion**

**Nano-FTIR spectroscopy on an LDPE-PS polymer blend.** Nano-FTIR and IR-sSNOM-imaging combine surface characterisation and optical spectroscopy at the nanoscale. In addition to topographical and mechanical phase data, local optical information of the surface of a sample is accessible.

**Figure 1** illustrates this by the example of a standard polymer blend sample with spherical low density polyethylene (LDPE) domains in a polystyrene (PS) matrix (Bruker PS-LDPE-12M). A white light nano-FTIR scan was conducted with incoming broadband infrared radiation of 2000-1000 cm$^{-1}$. Apart from topographic ($z$, Figure 1a) and mechanical phase ($\varphi_M$, Figure 1b) scan data, an optical NF-amplitude signal Figure 1c, deconvoluted at the second order of the tapping frequency, was obtained. Clear phase and optical contrasts are visible, in which phase shifts represent differences in energy dissipation between tip and sample, while a higher NF-amplitude indicates a stronger local reflection. Spectral differences between the domains become visible by interferometric detection of the optical trace as in the line scan across an LDPE domain in (Figure 1d). We recorded NF-phase spectra with a spatial resolution of 20 nm along a distance of 2 µm. Sharp changes occur between the domains and their optical signatures. These differences can be understood by comparing with point spectra recorded inside the spherical LDPE domain and the surrounding PS matrix with the same incident radiation of 2000-1000 cm$^{-1}$. Plots in Figure 1e show the relevant spectral region between 1700 and 1300 cm$^{-1}$. The positions of characteristic peaks for LDPE at 1460 cm$^{-1}$ (CH$_2$-bend) and for PS at 1601 cm$^{-1}$ (aromatic ring stretch), 1486 cm$^{-1}$ (aromatic ring stretch), 1445 (CH$_2$-bend) deviate only slightly from reference spectra in our library and literature,[11, 12] 7 cm$^{-1}$ at the most. The transition between the materials is almost immediate and occurs within one to two intermediate steps (Figure 1f, g). Changes in the relative intensity of the peak at 1486 cm$^{-1}$ are accompanied by a shift of the stronger peak from 1445 cm$^{-1}$ in PS to 1460 cm$^{-1}$ in LDPE. The highly resolved transition in the absorption indicates the existence of a thin phase with both signals present.

**Imaging of domain distributions.** Another possibility to gather chemical information on a nanoscale is IR-sSNOM imaging of the sample surface at defined mid-IR wavenumbers. Pseudo-heterodyne deconvolution of the optical signal at higher orders of the resonance frequency of the cantilever reveals clear NF-phase contrasts close to vibrational bands observed in the nano-FTIR mode. **Figure 2** is a presentation of optical amplitude and phase images of the LDPE-PS standard sample, recorded at different incident wavelengths. Close to the $CH_2$-bend vibration of LDPE, at 1467 cm$^{-1}$, the spherical LDPE domains are characterized by low NF-amplitudes (reflection) and high NF-phase (absorption) intensities. The image inverts when moving the excitation to 1640 cm$^{-1}$, close to the aromatic ring stretch vibration band of PS, while at 1710 cm$^{-1}$ neither compound absorbs strongly so that the contrast is weak. With measurement times of around three to ten minutes per image this technique allows for a fast and clear imaging of phase distributions in multicomponent polymer materials. Important to mention is that the original condition of the soft matter sample is not affected by the imaging process as the measurements are based on standard tapping-mode AFM technology.

**Library search.** In **Figure 3**, NF-amplitude and -phase spectra of polylactic acid (PLA) are plotted together with the real and imaginary parts of the data, which result from deconvolution at $n$ = 2 of the tapping frequency. We measured point spectra at three different spots on the sample and averaged these into one. Each measurement took 6:45 min, while earlier work reported more than 16 min.[2] We applied a minimum smoothing by five point moving averaging the data. Due to the lower laser intensity towards the rims of their spectral range, fluctuations in the background are more prominent there and only data between 1800 and 1050 cm$^{-1}$ was considered. As mentioned earlier, amplitude spectra and real part are related to the reflection at the nanofocus, while phase and imaginary part carry the information about the local infrared absorption of the sample. Confirming previous studies, phase spectra and imaginary part show slight differences in peak position and relative peak intensity (Figure 3a).[7] A close match occurs in direct comparison between the imaginary part spectrum and the ATR-IR reference of the same sample (Figure 3b). This underlines the suitability of imaginary part spectra or nano-FTIR absorption for library searches.

The applied broadband laser contains five separately accessible spectral ranges for irradiation of the sample. These are between 610-1400 cm$^{-1}$ (A), 700-1720 cm$^{-1}$ (B), 1000-2000 cm$^{-1}$ (C), 1200-2200 cm$^{-1}$ (D) and 1450-2200 cm$^{-1}$ (E). Based on the individual intensity profiles of the excitation spectra, the width of finally obtained sample spectra is about 700-800 cm$^{-1}$ per range. The majority of relevant signals which differentiate carbon based polymers within the fingerprint region lies within range C with output spectra between 1800 and 1050 cm$^{-1}$ (C=O stretch, aromatic stretch, CH$_2$-bend, C-N stretch and C-O stretch vibrations), so that this range is best suited for a straightforward general differentiation between polymers at the nanoscale. The intensity of infrared bands depends on the magnitude of the change in the dipole moment due to the vibration or rather their oscillator strength. Functional groups with carbon and oxygen or nitrogen heteroatoms as in ester groups of PLA often give rise to higher signal intensities than those containing only carbon and hydrogen. For samples with strong signals in the regions towards 1800 and 1100 cm$^{-1}$, the range can be broad as in the case of PLA. With polymers only containing C-H-vibrations and of these very few such as polyethylene (PE) fluctuations of the signal are even more noticeable in the regions of lower laser intensity, which effectively reduces the range of analysable spectral data. For this reason, the region between 1700 and 1300 cm$^{-1}$ was chosen for a general library search identification of some of the most ubiquitous polymers. The range covers characteristic aromatic and bending bands of C-H groups that are relevant for identification.

In order to evaluate the chance of differentiation between polymers within this range, we carried out a hierarchical cluster analysis based on our open access polymer reference library that is applied in the identification of microplastics.[13] The library contains 326 spectra of common commercial polymers and co-polymers as well as natural polymers such as cellulose. Similar to the adaptable database design[13] the Hellinger distance of the spectra was calculated with *PRIMER 6* software followed by hierarchical cluster analysis. We identified 112 differentiable clusters in the spectral range of 1700-1300 cm$^{-1}$ (for details see experimental section and Figure S1). This number is very close to the 107 clusters obtained in the same way for the wider range validated and applied in microplastics analysis (3600-1250 cm$^{-1}$). The analysis predicts in both cases that some relevant and spectrally similar compounds cannot be distinguished and thus belong to a common cluster. Examples for the clusters are: high + low density polyethylene, polyether/ polyester polyurethane, polyester including polyethylene terephthalate, polyalkyl methacrylates, and polystyrene + styrene acrylonitrile. A dendrogram with the results is shown in **Figure S1**. These results indicate a comparable identification success by near- and far field spectroscopy, with polymer types as a reliable result rather than exact compounds.

As mentioned earlier, the intensity of the irradiating laser is reduced towards the rims of the range. To assess which polymer spectra are affected to which degree, we further analysed the obtained spectral data within the range of 1800-1300 $cm^{-1}$, including C=O bands, and in the widest range of 1800-1070 $cm^{-1}$. Subjects of our examination were the following polymers: polylactic acid (PLA), polyamide (PA), polyethylene (PE), polystyrene (PS), polyvinyl chloride (PVC), polypropylene (PP), polyethylene terephthalate (PET), styrene acrylonitrile (SAN), polymethyl methacrylate (PMMA) and a polyether urethane (PEUR).

We carried out library searches with *OPUS 7.5 (©Bruker Optik GmbH)* software and our aforementioned library. Applying the "standard" search algorithm based on peak position, relative intensity and half-width of the peaks was not successful. It frequently led to low similarities between measured and returned reference spectra indicated by "hit qualities" (HQs; 0-1000, 1000 = identical spectra) and misidentification of the polymer in agreement with an earlier study on transmission FTIR spectra.[14] Thus, we applied a routine with search parameters set to vector normalization of the spectra and comparison of the first derivative, analogously to procedures validated earlier[14] (for exemplary spectra and library search results see **Figure S2-S8**). All polymer identifications were correct, only in one case another compound within the same cluster group was returned. The hit qualities as listed in **Table 1** are lower than those expected with larger scale FTIR but it has to be taken into account how extremely small the probed volume in the nanofocus is (sphere with the diameter of the cantilever tip, approx. $5*10^{-24}$ $m^3$). Moreover, three factors have a stronger impact than in other, larger scale methods: individual probe-sample interactions, small-scale changes on the surface or unknown coatings and environmental conditions. Electrostatic charging on the surface of polymers or unknown coatings may influence the interaction with the probing tip, a known phenomenon in atomic force microscopy, which may affect NF-interactions and spectra. Multiple point spectra from different spots should be

**Table 1:** Polymer types, their formula unit and hit qualities (HQ) for database comparison in the range of 1700-1300 cm$^{-1}$ and 1800-1300 cm$^{-1}$ without and with additional concave rubberband (RB) correction (and exclusion of water bands between 1600 and 1500 cm$^{-1}$) and over the complete accessible range 1800-1070 cm$^{-1}$ with additional rubber band correction.

| Polymer | Formula unit | 1700-1300 cm$^{-1}$ | | | 1800-1300 cm$^{-1}$ | | | 1800-1070 cm$^{-1}$ |
|---|---|---|---|---|---|---|---|---|
| | | HQ | HQ RB | H$_2$O excl. | HQ | HQ RB | H$_2$O excl. | HQ RB |
| PP | 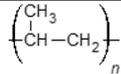 | 661 | 693 | 717 | 641 | 655 | 675 | 601 |
| PA | 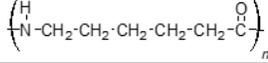 | 797 | 797 | * | 787 | 788 | * | 697 |
| PVC | 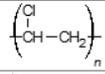 | 572 | 705* | * | 529 | 579 | * | 492 |
| PLA | 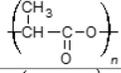 | 754 | 768 | 791 | 709 | 710 | 711 | 614 |
| PE | 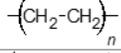 | 530 | 591 | 648 | 505 | 572 | 623 | 514 |
| PS | 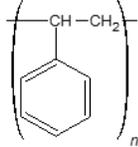 | 427 | 441 | 469** | 406 | 423 | 446** | 333 |
| SAN | 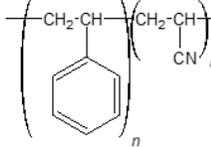 | 606 | 615 | 632 | 592 | 610 | 626 | 544 |
| PET | 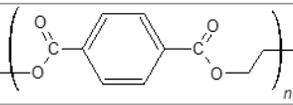 | 501 | 542 | 551 | 770 | 772 | 776 | 716 |
| PMMA | 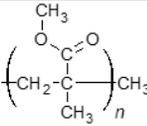 | 524 | 592 | 621 | 779*** | 801*** | 803*** | 739 |
| PEUR | 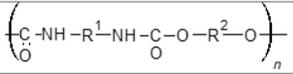 | 465 | 463 | * | 492 | 494 | * | 447 |

*Water bands were not excluded due to the presence of sample signals in the region. **The boundaries for water exclusion were set to 1584-1519 cm$^{-1}$. *** assigned to PB(utyl)MA within the same cluster of spectra.

averaged. As measurements are carried out under ambient conditions, water vapour may be visible in the detected spectra. The latter can be addressed by excluding respective regions from the analysed range.

PS exhibits the lowest HQ, which fits with the observation that this polymer is more difficult to detect by near-field spectroscopy in the relevant region compared to others.[10] PA and PLA on the other hand, are identified with HQs well above the limit of 700 for the classification as high quality data applied in far-field spectroscopy.[15] The intensity of recorded IR-spectra depends on the (local) refractive index of the sample as well as the oscillator strength of the respective vibration. In the case of nano-FTIR, which relies on scattering, the refractive index of the sample determines the efficiency of the backscattering process and thus the overall detected signal, so that higher intensities are expected for polymers with a high refractive index (> 1.5). The example of PLA with a low refractive index of 1.4 but strong characteristic vibrations,[16] on the other hand, emphasizes the role of bands with high oscillator strength in the identification of polymer samples at a small scale. PVC usually contains a large amount of additives, which in addition to water vapour may explain the wide signal above 1500 cm$^{-1}$. Despite its strong spectral similarity to PS, SAN is recognized as such, PEUR is identified as a polyurethane but a more specific assignment is not possible. The reason for this and the low HQs may be ageing of the material. For compounds with carbonyl groups it is beneficial to widen the range of analysis to 1800-1300 cm$^{-1}$, while added noise reduced the accuracy of all other identifications (see supplementary Figures). All HQs increased, however, when further adding a concave rubberband correction to the routine in order to create a flattened baseline. For weaker absorbers without spectral information in the same region, hit qualities continued to increase when excluding bands at 1600-1500 cm$^{-1}$, which result from water vapour in the atmosphere (compare HQ PE with and without the exclusion of water signals between 1600 and 1500 cm$^{-1}$).

Broadening the analysed range of the data increased the contribution of noise in form of additional peaks to most spectra and may cause irregularities reducing the HQs. A clear benefit may however be the ability to detect carbonyl groups within the same range. Best results for PMMA, PET and PEUR were obtained with 1800-1300 cm$^{-1}$. PMMA is assigned to polybutyl methacrylate in this region, which belongs to the same cluster of indistinguishable spectra. When requiring more spectral information than available in the aforementioned ranges, spectra from three different incident ranges can be combined into one from 2000 to 670 cm$^{-1}$. In the spectrum of PS it becomes clear that the strongest signal is the one at 697 cm$^{-1}$ (**Figure 4**). For PE-type polymers additional signals in the lower wavenumber range are comparatively weak and do not significantly alter the detection accuracy with HQs for the full region of 454 and 506 with rubber band correction. A significant gain occurs with PS, reaching HQs of 698 before and 700 after rubber band correction. PA is less accurately identified than in the smaller range as mostly noise and weak signals are added to the spectrum. This results in HQs of 592 before and 601 after rubber band correction. Between 1300 and 1100 cm$^{-1}$ a broader feature underlies the signals. This feature is caused by a background effect that cannot be easily separated from the signals of the sample. Reducing the analysed range to 1700-1300 cm$^{-1}$ + 800-670 cm$^{-1}$, removes noisier parts of the spectra and results in further improved hit qualities, PS: 717, PE: 534 and PA: 648. In conclusion, measuring a wider range is only useful if important signals are captured, otherwise increased noise or water vapour peaks disturb the analysis.

Depending on the application and diversity of possible samples, we thus propose to apply different strategies. For general identification tasks and higher throughput a range of 1700-1300 or 1800-1300 cm$^{-1}$ can be applied in combination with post-processing of the data. Polymers with weak signals such as PE are more easily identified when the spectral range matches their most intense signals and is kept as short as possible. Combining data from more than one range requires time but is justified when main peaks are suspected outside the range of general identification.

**Experimental**

**Polymer samples.** A PS-LDPE-12M reference sample was obtained from Bruker. Commercial polymer foil samples were of different sources: PP (400 µm, Dr. Dietrich Müller GmbH), PA (1000 µm, Dr. Dietrich Müller GmbH), PLA (300 µm, Folienwerk Wolfen GmbH), PE (high density, 99 µm; low density, 99 µm Orbita Film GmbH), PS (Ergo.fol, 190 µm, Norflex GmbH), PVC (210 µm, Leitz), styrene acrylonitrile SAN (Ergo.fol, 90 µm, Norflex GmbH), PMMA (pellet, cut, Bayreuth University), PEUR (LPT 4802 T 050 natural, Bayer), PET (175 µm, Pütz GmbH + Co. Folien KG). The samples were cut to pieces of < 1 cm$^{-2}$ and cleaned by wiping with HPLC grade ethanol (Merck) on a lint-free paper and blowing off dust and fibre fragments.

**Spectroscopy**

All experiments were carried out in ambient atmosphere with a neaspec neaSNOM system coupled with two different infrared light sources. Measurements were conducted in tapping mode with metal-coated cantilevers at resonance frequencies of 50-250 kHz. Scan data was levelled and line corrected with *Gwyddion 2.49*.[17]

**Nano FTIR.** The laser source for spectroscopic measurements was a neaspec nano-FTIR mid IR super continuum laser with ranges of 610-1400 cm$^{-1}$(A), 700-1720 cm$^{-1}$ (B), 1000-2000 cm$^{-1}$ (C), 1200-2200 cm$^{-1}$ (D) and 1450-2200 cm$^{-1}$ (E), a bandwidth of approximately 700-800 cm$^{-1}$ and tuneable output in the spectral range of 2000-670 cm$^{-1}$ with powers of 0.05-1.0 mW.

The applied settings were: Interferometer centre 400 µm, interferometer distance 500 µm (total available path length: 800 µm), 10 cm-1 spectral resolution, 2048 pixels and an integration time of 10 ms per pixel with 20 scans averaged per point. The set point was 80%, while the free tapping amplitude was set to 100 nm, resulting in an amplitude of approximately 70 nm in approached tapping state. After the measurements, the spectra were normalized to a reference measured on a Si wafer, phase corrected and underwent a manual five point moving average treatment within the instrument software. Spectra obtained from combining multiple range (A, B, D) spectra were manually joined at points of spectral overlap of the phase component with neaplotter software.

IR-sSNOM imaging. The source of monochromatic irradiation was a Daylight solutions MIRcat laser with tuneable wavenumbers between 1872 and 903 cm-1. Output powers were tuned to approximately 1 mW. The pixel time for the scans was 7 ms.

**Data analysis**

Library search identification of polymer films. For each sample, three spots were measured with the same settings and the obtained nano-FTIR spectra were averaged. Analogous to established procedures for ATR-IR and µ-FTIR based identification of polymers, the resulting spectrum was then compared to the group's open access database of polymer spectra for automated analysis with the software OPUS (OPUS 7.5, Bruker Optik GmbH).13 In case of the wider spectral ranges our in-house database for manual identification was applied (the range of the database for automated analysis is 3600-1250 cm 1). The library search was carried out for spectral data based on vector normalization and the first derivative of the curve.13, 14 Hit qualities were determined without and with individual concave rubber band background correction (64 baseline points, number of iterations: PA 1700/1800-1070 cm-1 = 1, 2000-670 cm-1 = 21; PLA = 5; PP = 10; PVC, PS = 21, PE = 51) and exclusion of water bands between 1600 and 1500 cm-1, as indicated.

Hierarchical cluster analysis. ATR reference spectra underwent a hierarchical cluster analysis using the Primer 6 software equipped with the Permanova+ package (PRIMER-E) in the range of 1700-1300 cm-1. For this, all spectra were offset corrected starting from 0. The data was normalised to percentage to exclude effects from different concentrations and varying contacts between diamond crystal and material during the ATR measurement. Prior to cluster analysis, the Hellinger distance of the different spectra was calculated.

**Conclusions**

Nano-FTIR and IR-sSNOM-imaging provide fast and detailed nanoscale characterisation and identification of polymer samples and blends through the combination of high-resolution microscopy and spectroscopy. Within a few minutes, the distribution of components and their phase boundaries can be analysed with resolutions down to 20 nm. Data obtained by nano-FTIR correctly identifies all the examined polymer samples with good to high quality results at moderate measurement times of less than seven minutes per spectrum and with a minimum post-processing routine. It is possible to differentiate between the most abundant polymer types comparable to conventional ATR-IR spectroscopy. The spectral range of only 1700-1300 cm-1 is sufficient for this. When higher accuracies are required, additional processing of the data such as rubber band baseline corrections may be applied or spectral ranges increased to further improve hit qualities and thus identification certainties. For materials with weak and few signals such as PE and PS, small ranges are beneficial. Discussions remain on whether and how to adjust thresholds for accepted HQ numbers for future identification tasks.

Our study proves the applicability of standard identification tools for unknown polymer samples to nano-FTIR data. Such a finding is of major interest, for example in the identification of unknown polymeric nanomaterials and raises high hopes for an unambiguous identification of nanostructured polymers and particles in environmental samples.

**Acknowledgements**

The authors thank Andreas Huber of neaspec for experimental support and fruitful discussions. The work was funded through the project "Size is important" (FIT 12317001) by the WTSH Business Development and Technology Transfer Corporation of Schleswig Holstein. SP and GG further thank the German Federal Ministry of Education and Research for grant 03F0734A (BASEMAN - Defining the baselines and standards for microplastics analyses in European waters, BMBF).

further thank the German Federal Ministry of Education and Research for grant 03F0734A (BASEMAN - Defining the baselines and standards for microplastics analyses in European waters, BMBF).

further thank the German Federal Ministry of Education and Research for grant 03F0734A (BASEMAN - Defining the baselines and standards for microplastics analyses in European waters, BMBF).

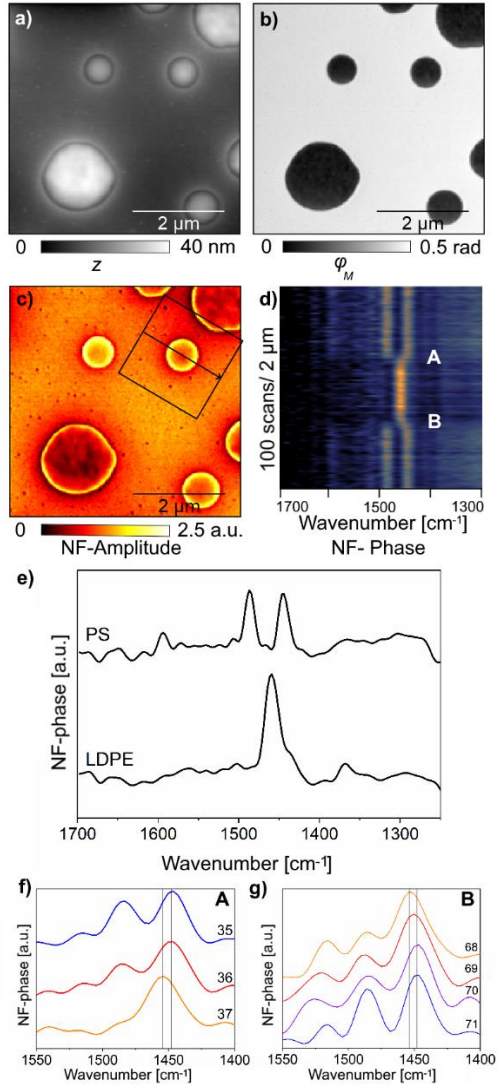

**Figure 1.** Nano-FTIR scans with a) topography ($z$), b) mechanical phase ($\varphi_M$) and c) NF-amplitude ($n$ = 2) signals of a standard polymer blend sample with spherical LDPE domains in a PS matrix. d) NF-phase ($n$ = 2) of a line-scan with a resolution of 20 nm through an LDPE domain. e) Nano-FTIR NF-phase point spectra of the different materials in the same spectral region (1700-1300 cm$^{-1}$). f) and g) zooms of spectra recorded in the upper (A) and lower (B) transition zones. The lower NF-phase peak shifts from 1445 cm$^{-1}$ to 1460 cm$^{-1}$ from PS to LDPE.

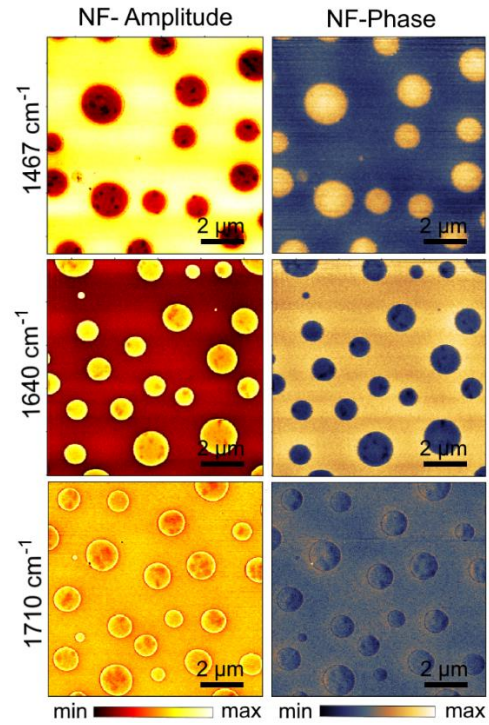

**Figure 2.** IR-sSNOM imaging of an LDPE-PS sample at different wavenumbers, close to the $CH_2$-bend/aromatic ring stretch vibrations of LDPE (1467 cm$^{-1}$), the aromatic ring stretch vibration of PS (1640 cm$^{-1}$) and far from resonances of the two (1710 cm$^{-1}$). Intensity distributions in amplitude and phase spectra demonstrate the switch from low reflection (amplitude) and high absorption (phase) to the opposite when changing from one material's resonance to the other's. All data are deconvoluted at $n$ = 2.

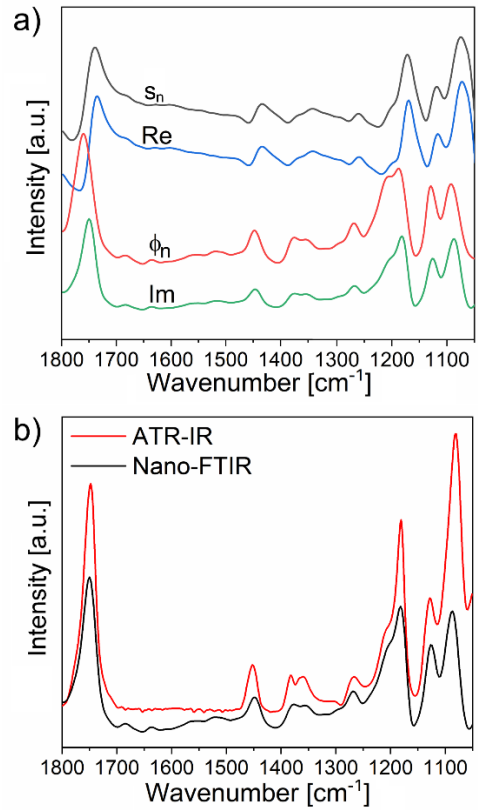

**Figure 3.** a) Amplitude ($s_n$), real part (Re), NF phase ($\varphi_n$) and imaginary part (Im) plots of NF spectra and b) comparison of Im, labelled nano-FTIR, and ATR-IR spectra of PLA bands observed in the nano-FTIR mode.

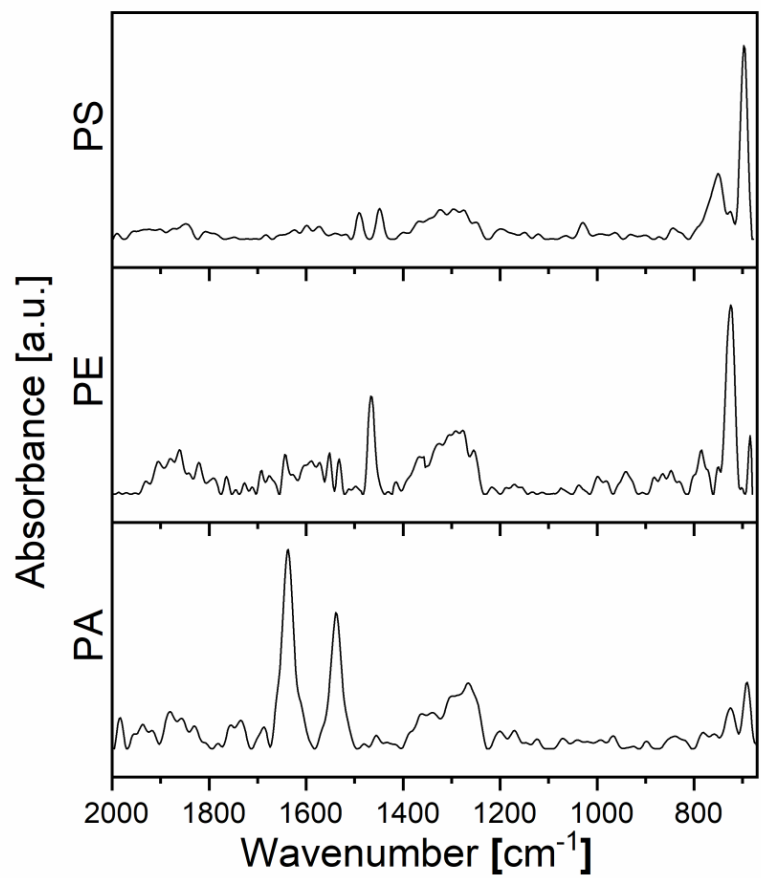

**Figure 4.** Combined range spectra between 2000 and 670 cm$^{-1}$ of PS, PE and PA, concave rubber band corrected.

**Supplementary Figures**

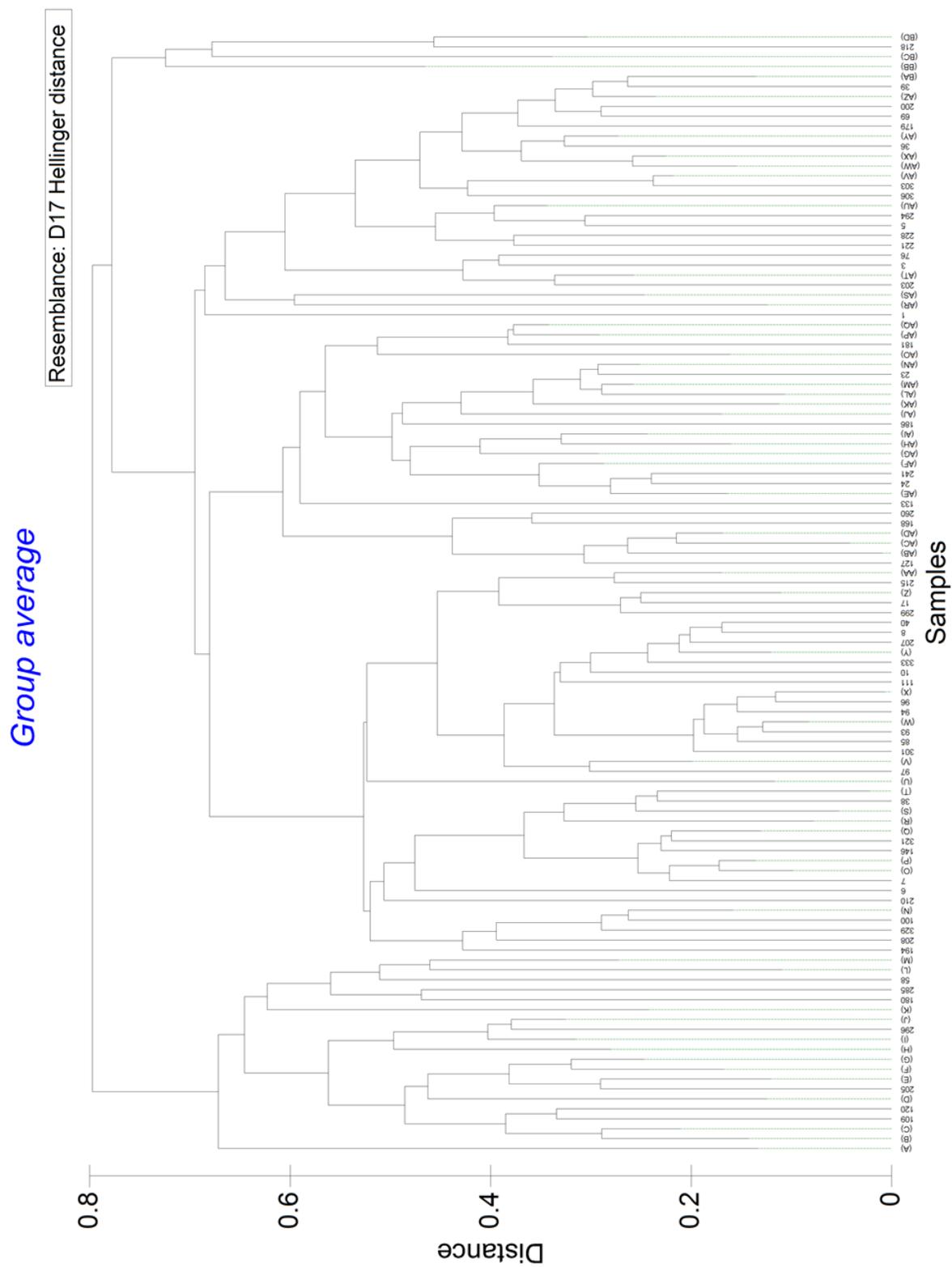

**Figure S1.** Dendrogram for cluster analysis of reference spectra in the range of 1700-1300 cm$^{-1}$. The numbers are representing the ID of the spectrum of the database available in reference 1. A:187+188 (PEG), B:117+272+273 (LDPE), C: 53+12+13+52+54+88+80+319+124+224+86+87+276+277+278+219+118+119+116+320+300+225+ 123+270 (PE), D:78+79 (EVA), E:154+131+63+83+84 (Bee wax), F:74+75 (EVAc), G:71+14+226 (PE wax), H:209+176+177 (PLA), I:178+183 (P1B,P4M1P), J:231+108+325+143+248+247+252+249+229+230+232+250 (PP), K:56+73 (PE-PP copolymer), L:206+244 (POM), M:220+222 (PE-chlorinated), N:55+60 (EPDM), O:330+332+331+334 (Coal), P:302+144+289 (Silk), Q:266+50+11+115+30+313+132+27+28+15+126+267+287+ 26+29+312+311+315+62+268+314+316+310+317 (Fur), R:149+281+151+152+279+148+147+156+155+160 +150+41+42+280+137+105+161+159+157+158, S:196+235 (PA), T:37+51 (Chitin), U:134+135 (Aramid), V:264+265 (Quartz), W:16+286 (Bee wax aged), X:104+106 (fiber natural jute/kapok), Y:245+246 (fiber poplar), Z:121+9+81 (fiber linen/algae), AA:130+129+184 (PAA,MVEMA), AB:153+243 (PVP), AC:238+239 (PUR 1), AD:258+259 (PEUR 1), AE:284+282+283+256+257 (PEUR 2), AF:91+92 (alkyd varnish 1), AG:211+212+236+237 (PCL/PUR), AH:2+182 (ABS/PDDPF), AI:141+142 (PEST 1), AJ:18+20+21 (PEST resin), AK:139+216+217+227+171+ 172+136+323+174+98+99 (PEST/PET 1), AL:274+275+44+45 (PEST 2), AM:173+189 (PET), AN:43+170+164+ 163+185 (PBT), AO:261+65+66 (PEST 3), AP:70+72 (EAA/EMA), AQ:192+324+190+191+25+193 (P(R)MA), AR:295+125+112+114 (alkyd varnish 2), AS:204+242 (PVF), AT:67+288 (Silicone rubber), AU:4+95+197+33+35+ 32+318 (CA), AV:145+328+223+297 (nitrile and natural rubber), AW:307+201+327+202+107+263 (PVC), AX:101+102+128 (hydroxypropyl-Ce), AY:309+89+90+47+48+49+103+308+113+31+269+34+82+122 (Ce), AZ:140+304+305 (Carboxylated PVC), BA:198+199+77+240 (PVA), BB:59+57+61 (rubber), BC:298+291+290+ 293+271+292+255+233+254+253+326+64+68 (PS), BD:175+138+165+166+214+ 213+322 (PC), BE:234+195+251 (P(P)SU).

**Library search results**

The following graphs show nano-FTIR spectra and library search reference spectra of polypropylene after different data treatment steps in red and blue, respectively. All spectra are maximized in absorbance units here for better visibility. The number in the field of compound name (*Substanzname*) is an indicator for the cluster of spectroscopically indistinguishable compounds that the reference spectrum is assigned to in automatic analyis.[1]

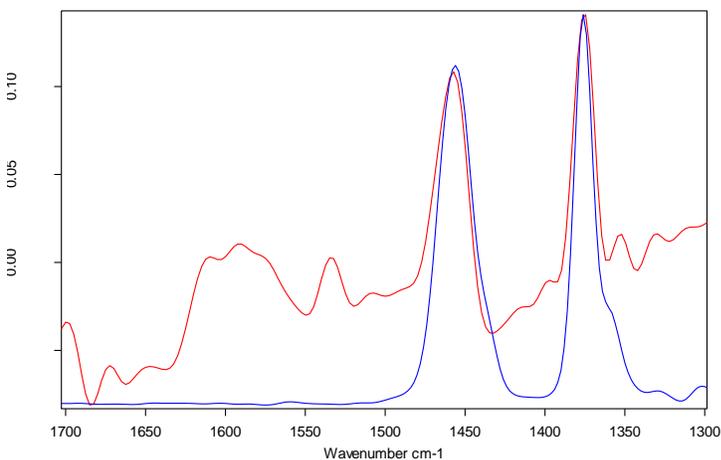

| Substanz | 4 |
| Kurzzeichen | PP |
| Hersteller | Dr. Dietrich Müller GmbH |
| Form (Pulver, Pellet, Folie, Stü | Folie |
| Eintrag Nr. | 101 |
| Bibliotheksname | BASEMAN_AUTOMATED.S01 |
| Bibliotheksbeschreibung | Finale Bibliothek für die automatische Ausw |
| Copyright | Sebastian Primpke |

| Color | Hit Quality | Compound name | CAS Number | Molecular formula | Molecular weight |
|---|---|---|---|---|---|
|  | 691 | 4 |  |  |  |

**Figure S2.** Polypropylene 1700-1300 cm$^{-1}$, uncorrected.

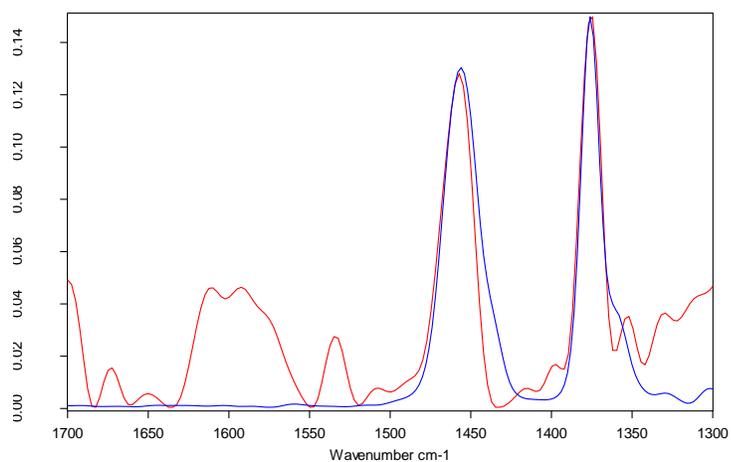

| Color | Hit Quality | Compound name | CAS Number | Molecular formula | Molecular weight |
|---|---|---|---|---|---|
|  | 698 | 4 |  |  |  |

**Figure S3.** Polypropylene, 1700-1300 cm$^{-1}$ with rubberband correction.

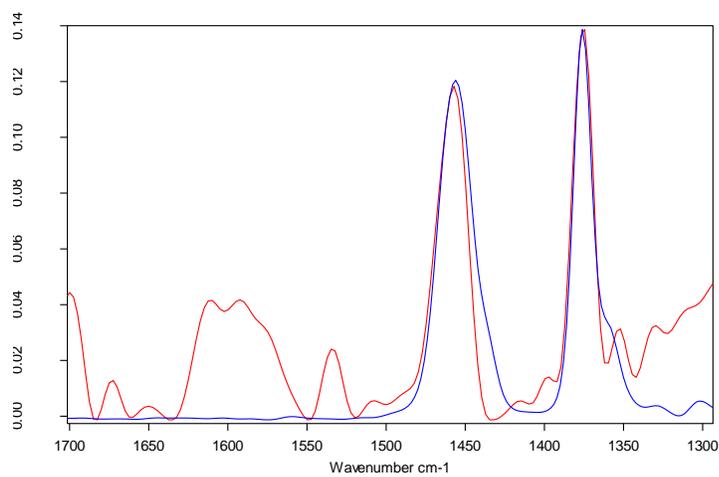

| Color | Hit Quality | Compound name | CAS Number | Molecular formula | Molecular weight |
|---|---|---|---|---|---|
|  | 724 | 4 |  |  |  |

**Figure S4.** Polypropylene, 1700-1300 cm$^{-1}$ with rubberband correction and exclusion of water (1600- 1500 cm$^{-1}$).

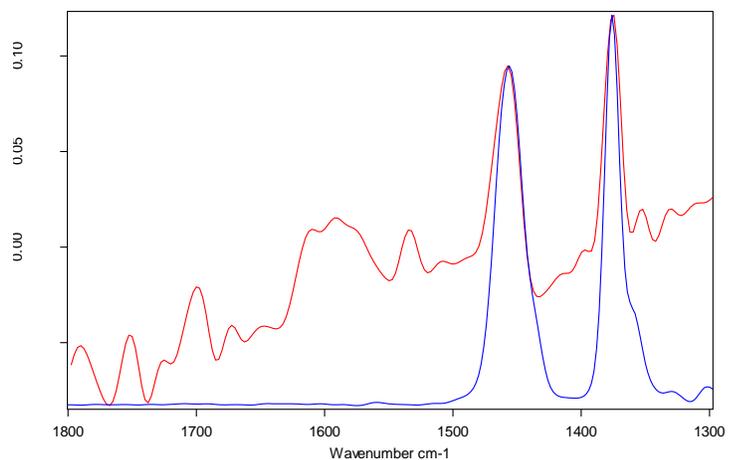

| Color | Hit Quality | Compound name | CAS Number | Molecular formula | Molecular weight |
|---|---|---|---|---|---|
| 🟦 | 641 | 4 | | | |

**Figure S5.** Polypropylene 1800-1300 cm$^{-1}$, uncorrected.

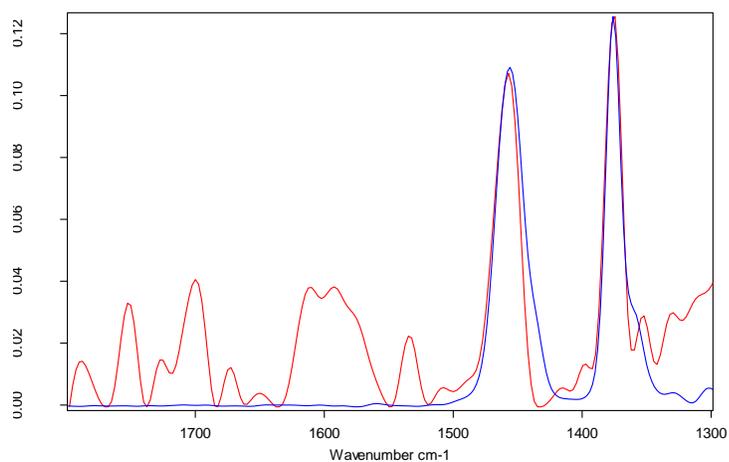

| Color | Hit Quality | Compound name | CAS Number | Molecular formula | Molecular weight |
|---|---|---|---|---|---|
| 🟦 | 655 | 4 | | | |

**Figure S6.** Polypropylene, 1800-1300 cm$^{-1}$ with rubberband correction.

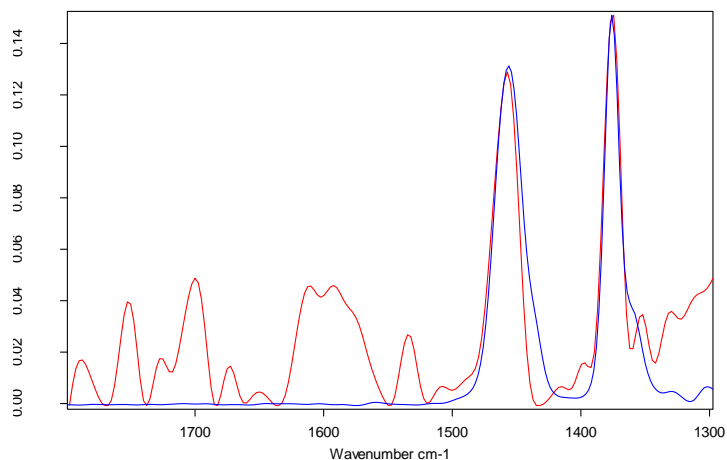

| Color | Hit Quality | Compound name | CAS Number | Molecular formula | Molecular weight |
|---|---|---|---|---|---|
|  | 675 | 4 |  |  |  |

**Figure S7.** Polypropylene, 1800-1300 cm$^{-1}$ with rubberband correction and exclusion of water (1600-1500 cm$^{-1}$).

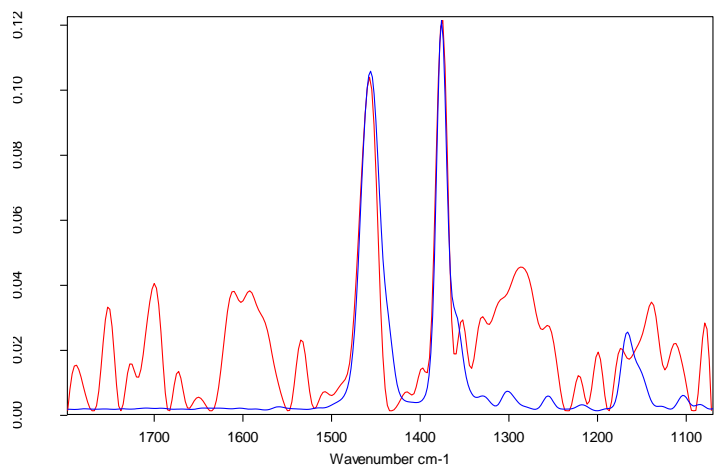

| Color | Hit Quality | Compound name | CAS Number | Molecular formula | Molecular weight |
|---|---|---|---|---|---|
|  | 601 | 4 |  |  |  |

**Figure S8.** Polypropylene, 1800-1070 cm$^{-1}$ with rubberband correction.